\documentclass[12pt]{article}

\begin{document}
\begin{center}
{\bf On the Hamiltonian Form of Generalized Dirac Equation for Fermions with Two Mass States}\\
\vspace{5mm}
 S. I. Kruglov \\
\vspace{5mm}
\textit{University of Toronto at Scarborough,\\ Physical and Environmental Sciences Department, \\
1265 Military Trail, Toronto, Ontario, Canada M1C 1A4}
\end{center}

\begin{abstract}
Dynamical and non-dynamical components of the 20-component wave
function are separated in the generalized Dirac equation of the
first order, describing fermions with spin 1/2 and two mass
states. After the exclusion of the non-dynamical components, we
obtain the Hamiltonian Form of equations. Minimal and non-minimal
electromagnetic interactions of particles are considered here.

\end{abstract}

\section{Introduction}

We continue to investigate the first order generalized Dirac
equation (FOGDE), describing fermions with spin 1/2 and two mass
states. This 20-component wave equation was obtained in
\cite{Kruglov} on the base of Barut's \cite{Barut} second order
equation describing particles with two mass states. Barut
suggested the second order wave equation for the unified
description of $e$, $\mu$ leptons. He treated this equation as an
effective equation for partly ``dressed" fermions using the
non-perturbative approach to quantum electrodynamics. Some
investigations of Barut's second order wave equation and FOGDE
were performed in \cite{Barut1}, \cite{Barut2}, \cite{Wilson},
\cite{Dvoeglazov}, \cite{Kruglov1}.

The purpose of this paper is to obtain the Hamiltonian Form of the
20-component wave equation of the first order.

The paper is organized as follows. In Sec. 2, we introduce the
generalized Dirac equation of the first order. The dynamical and
non-dynamical components of the 20-component wave function are
separated, and quantum-mechanical Hamiltonian is derived in Sec.
3. In Sec. 4, we make a conclusion. In Appendix, we give some
useful matrices entering the Hamiltonian. The system of units
$\hbar =c=1$ is chosen, Latin letters run 1, 2, 3, and Greek
letters run 1, 2, 3, 4, and notations as in \cite{Ahieser} are
used.

\section{Field Equation of the First Order}

The Barut second order field equation describing spin-1/2 and two
mass states of particles may be rewritten as \cite{Kruglov}:
\begin{equation}
\left(\gamma_\mu\partial_\mu- \frac{a}{m}\partial_\mu^2 +
m\right)\psi(x)=0 , \label{1}
\end{equation}
where $\partial_\mu =\partial/\partial x_\mu =(\partial/\partial
x_m,\partial/\partial (it))$, $\psi (x)$ is a Dirac spinor, $m$ is
a parameter with the dimension of the mass, and $a$ is a massless
parameter. We imply a summation over repeated indices. The
commutation relations $\gamma_\mu \gamma_\nu +\gamma_\nu
\gamma_\mu =2\delta_{\mu\nu}$ are valid for the Dirac matrices.
Masses of fermions are given by
\begin{equation}
m_1=\pm m\left(\frac{1 - \sqrt{4a+1}}{2a}\right) ,~~m_2=\pm
m\left(\frac{1 +\sqrt{4a+1}}{2a}\right) .
 \label{2}
\end{equation}
Signs in Eq. (2) should be chosen to have positive values of
$m_1$, $m_2$.

Eq. (1) can be represented in the first order form \cite{Kruglov}:
\begin{equation}
\left( \alpha _\mu \partial _\mu +m\right) \Psi (x)=0 , \label{3}
\end{equation}
where the 20-dimensional wave function $\Psi (x)$ and
$20\times20$-matrices $\alpha _\mu$ are
\begin{equation}
\Psi (x)=\left\{ \psi _A(x)\right\} =\left(
\begin{array}{c}
\psi (x)\\
\psi _\mu (x)
\end{array}
\right) \hspace{0.5in}(\psi_\mu (x)=-\frac{1}{m}\partial_\mu \psi
(x)) , \label{4}
\end{equation}
\begin{equation}
\alpha _\mu =\left(\varepsilon ^{\mu,0 }+ a\varepsilon ^{0,\mu
}\right)\otimes I_4 + \varepsilon^{0,0}\otimes\gamma_\mu
.\label{5}
\end{equation}
The $I_4$ is a unit $4\times4$-matrix, and $\otimes$ is a direct
product of matrices. The elements of the entire algebra obey
equations as follows (see, for example, \cite{Kruglov2}):
\begin{equation}
\left( \varepsilon ^{M,N}\right) _{AB}=\delta _{MA}\delta _{NB},
\hspace{0.5in}\varepsilon ^{M,A}\varepsilon ^{B,N}=\delta
_{AB}\varepsilon ^{M,N}, \label{6}
\end{equation}
where $A,B,M,N=0,1,2,3,4$.

After introducing the minimal electromagnetic interaction by the
substitution $\partial _\mu \rightarrow D_\mu =\partial _\mu
-ieA_\mu $ ($A_\mu $ is the four-vector potential of the
electromagnetic field), and the non-minimal interaction with the
electromagnetic field by adding two parameters $\kappa _1$,
$\kappa _2$, we come \cite{Kruglov} to the matrix equation:
\begin{equation}
\biggl [\alpha _\mu D_\mu +\frac i2\left( \kappa _0P_0+\kappa _1
P_1\right) \alpha _{\mu \nu }\mathcal{F}_{\mu \nu }+m\biggr ]\Psi
(x)=0  ,\label{7}
\end{equation}
where $P_0=\varepsilon ^{0,0}\otimes I_4$, $P_1=\varepsilon ^{\mu
,\mu }\otimes I_4$ are the projection operators, $P_0^2=P_0$,
$P_1^2=P_1$, $P_0+P_1=1$, and $\alpha _{\mu \nu }=\alpha _\mu
\alpha _\nu -\alpha _\nu \alpha _\mu$. Parameters $\kappa_0$ and
$\kappa_1$ characterize fermion anomalous electromagnetic
interactions.

The tensor form of Eq. (7) is given by
\begin{equation}
\left( \gamma_\nu D_\nu +i\kappa_0 \gamma_\mu \gamma_\nu
\mathcal{F}_{\mu \nu} + m\right)\psi (x)+\left(aD_\mu
+i\kappa_0\gamma_\nu\mathcal{F}_{\nu \mu}\right)\psi_\mu (x)=0 ,
 \label{8}
\end{equation}
\begin{equation}
\left(D _\mu +i\kappa_1\gamma_\nu\mathcal{F}_{\mu \nu}\right) \psi
(x)+ \left(m\delta_{\mu\nu}+i\kappa_1 a\mathcal{F}_{\mu
\nu}\right)\psi_\nu (x)=0 ,
 \label{9}
\end{equation}
where $\mathcal{F}_{\mu \nu }=\partial _\mu A_\nu -\partial _\nu
A_\mu $ is the strength of the electromagnetic field. Eq. (8), (9)
represent the system of equations for Dirac spinor $\psi (x)$ and
vector-spinor $\psi_\nu (x)$ interacting with electromagnetic
fields.

\section{Quantum-Mechanical Hamiltonian}

In order to obtain the quantum-mechanical Hamiltonian, we rewrite
Eq. (7) as follows:
\[ i\alpha _4\partial _t\Psi (x)=\biggl [\alpha
_aD_a+m+eA_0\alpha _4+
\]
\vspace{-7mm}
\begin{equation} \label{10}
\end{equation}
\vspace{-7mm}
\[
+\frac{i}{2}\left( \kappa _0 P_0+\kappa _1P_1 \right) \alpha _{\mu
\nu }\mathcal{F}_{\mu \nu }\biggr ]\Psi (x).
\]
One can verify with the help of Eq. (6) that the matrix $\alpha
_4$ obeys the matrix equation
\begin{equation}
 \alpha _4^4-(1+2a)\alpha_4^2 +a^2 \Lambda =0 ,
\label{11}
\end{equation}
where $\Lambda$:
\begin{equation}
 \Lambda =\left(\varepsilon ^{0,0 }+ \varepsilon ^{4,4}
 \right)\otimes I_4 ,
 \label{12}
\end{equation}
 is the projection operator, $\Lambda^2=\Lambda$.
It should be noted that the matrix $\Lambda$ can be considered as
the unit matrix in the $8$-dimensional sub-space of the wave
function \cite{Kruglov}. The operator $\Lambda$, acting on the
wave function $\Psi(x)$, extracts the dynamical components $\Phi
(x)=\Lambda \Psi(x)$. We may separate\footnote{In the work
\cite{Kruglov}, the dynamical and non-dynamical components of the
wave function were not separated.} the dynamical and non-dynamical
components of the wave function $\Psi (x)$ by introducing the
second projection operator:
\begin{equation}
\Pi =1-\Lambda = \varepsilon ^{m,m }\otimes I_4   , \label{13}
\end{equation}
so that $\Pi^2=\Pi$. This operator defines non-dynamical
components $\Omega=\Pi \Psi (x)$. Multiplying Eq. (10) by the
matrix
 \[
\frac{(1+2a)}{a^2}\alpha_4-\frac{\alpha_4^3}{a^2}=
\left(\varepsilon ^{0,4}+\frac{1}{a}\varepsilon
^{4,0}\right)\otimes I_4 -\frac{1}{a}\varepsilon ^{4,4}\otimes
\gamma_4 ,
\]
and taking into consideration Eq. (11), we obtain the equation as
follows:
\begin{equation}
i\partial _t \Phi (x)=eA_0 \Phi (x) +
\left[\frac{(1+2a)}{a^2}\alpha_4-\frac{\alpha_4^3}{a^2}\right]\biggl
[\alpha_a D_a+m +K\biggr ]\Psi (x), \label{14}
\end{equation}
where
\[
K=\frac{i}{2}\left( \kappa _0 P_0+\kappa _1P_1 \right) \alpha
_{\mu \nu }\mathcal{F}_{\mu \nu }
\]
\vspace{-7mm}
\begin{equation} \label{15}
\end{equation}
\vspace{-7mm}
\[
=i\mathcal{F}_{\mu \nu }\left[\kappa _0\left(\varepsilon
^{0,0}\otimes \gamma_\mu \gamma_\nu+\varepsilon ^{0,\nu}\otimes
\gamma_\mu\right)+\kappa _1\left(\varepsilon ^{\mu,0}\otimes
\gamma_\nu+a\varepsilon ^{\mu,\nu}\otimes I_4\right)\right] .
\]
It should be mentioned that because $\Lambda+\Pi=1$, the equality
$\Psi(x)=\Phi(x)+\Omega (x)$ is valid. To eliminate the
non-dynamical components $\Omega (x)$ from Eq. (14), we multiply
Eq. (10) by the matrix $\Pi$, and using the equality $\Pi \alpha_4
=0$, we obtain
\begin{equation}
\Pi \left(\alpha_a D_a +K\right)\left(\Phi(x)+\Omega (x)\right)+
m\Omega (x)=0 . \label{16}
\end{equation}
With the help of equation $\Pi \alpha_a \Pi=0$, one may find from
Eq. (16), the expression as follows:
\begin{equation}
\Omega(x)=-\left(m+\Pi K\right)^{-1}\Pi\left(\alpha_a D_a +
K\right)\Phi(x) . \label{17}
\end{equation}
With the aid of Eq. (17), Eq. (14) takes the form
\begin{equation}
i\partial _t\Phi (x)=\mathcal{H}\Phi (x) , \label{18}
\end{equation}
\[
\mathcal{H}=eA_0 + \left[\frac{(1+2a)}{a^2}\alpha_4
-\frac{\alpha_4^3}{a^2}\right]\biggl [\alpha_a D_a+m +K\biggr ]
\]
\vspace{-7mm}
\begin{equation} \label{19}
\end{equation}
\vspace{-7mm}
\[
\times\left[1-\left(m+\Pi K\right)^{-1}\Pi\left(\alpha_b D_b +
K\right)\right] ,
\]
Eq. (18) represents the Hamiltonian form of the equation for
$8$-component wave function $\Phi(x)$. It is obvious that for the
relativistic description of fermionic fields, possessing two mass
states, it is necessary to have $8$-component wave function (two
bispinors). The Hamiltonian (19) can be simplified by using
products of matrices given in Appendix.

Now we consider the particular case of fermions minimally
interacting with electromagnetic fields, $\kappa_0=\kappa_1=0$,
$K=0$. In this case, Eq. (18) becomes
\[
i\partial _t\Phi (x)= \biggl[eA_0 + \frac{m}{a}\left(a \varepsilon
^{0,4}\otimes I_4 +\varepsilon ^{4,0}\otimes I_4-\varepsilon
^{4,4}\otimes \gamma_4\right)
\]
\vspace{-7mm}
\begin{equation} \label{20}
\end{equation}
\vspace{-7mm}
\[
+\frac{1}{a}\left(\varepsilon ^{4,0}\otimes \gamma_m\right) D_m
-\frac{1}{m}\left(\varepsilon ^{4,0}\otimes I_4
\right)D_m^2\biggr]\Phi (x) .
\]
In component form, Eq. (20) is given by the system of equations
\[
i\partial _t\psi (x)= eA_0 \psi (x)+  m\psi_4 (x) ,
\]
\vspace{-7mm}
\begin{equation} \label{21}
\end{equation}
\vspace{-7mm}
\[
i\partial _t\psi_4 (x)=\left(eA_0-\frac{m}{a}\gamma_4\right)
\psi_4 (x)+\left(\frac{m}{a}+\frac{1}{a}\gamma_m
D_m-\frac{1}{m}D_m^2 \right)\psi (x) .
\]
Eq. (21) can also be obtained from Eq. (8), (9), at
$\kappa_0=\kappa_1=0$, after the exclusion of non-dynamical
(auxiliary) components $\psi_m (x)=-(1/m)D_m \psi (x)$. So, only
components with time derivatives enter Eq. (21) and Eq. (18).

\section{Conclusion}

We have analyzed the 20-component matrix equation of the first
order, describing fermions with spin 1/2 and two mass states which
is convenient for different applications. There are two parameters
characterizing non-minimal electromagnetic interactions of
fermions including the interaction of the anomalous magnetic
moment of particles. The Hamiltonian form of the equation was
obtained, and it was shown that the wave function, entering the
Hamiltonian equation, contains $8$ components what is necessary
for describing fermionic field with two mass states in the
formalism of the first order. The Hamiltonian (19) can be used for
a consideration of the non-relativistic limit which is convenient
for the physical interpretation of constants $\kappa_0$,
$\kappa_1$ introduced. This can be done with the help of the Foldy
- Wouthuysen procedure \cite{Foldy}.

The approach developed may be applied for a consideration of two
families of leptons or quarks, but this requires further
investigations.

\section{Appendix}

With the help of Eq. (6), one can obtain expressions as follows:
\begin{equation}
\left[\frac{(1+2a)}{a^2}\alpha_4-\frac{\alpha_4^3}{a^2}\right]\alpha_m
D_m =\left(\frac{1}{a}\varepsilon ^{4,0}\otimes \gamma_m +
\varepsilon ^{4,m} \otimes I_4 \right)D_m ,\label{22}
\end{equation}
\begin{equation}
\Pi \alpha_m D_m  = \left(\varepsilon ^{m,0}\otimes I_4 \right)D_m
,
 \label{23}
\end{equation}
\begin{equation}
\Pi K= i\kappa _1\mathcal{F}_{m \nu }\left(\varepsilon
^{m,0}\otimes \gamma_\nu+a\varepsilon ^{m,\nu}\otimes I_4\right) ,
 \label{24}
\end{equation}
\[
\left[\frac{(1+2a)}{a^2}\alpha_4-\frac{\alpha_4^3}{a^2}\right] K =
i\frac{\kappa_0}{a}\mathcal{F}_{\mu \nu }\left(\varepsilon
^{4,0}\otimes \gamma_\mu\gamma_\nu +\varepsilon ^{4,\nu}\otimes
\gamma_\mu\right)
\]
\vspace{-7mm}
\begin{equation} \label{25}
\end{equation}
\vspace{-7mm}
\[
+i\kappa_1\mathcal{F}_{4 \nu }\left(\varepsilon ^{0,0}\otimes
\gamma_\nu -\frac{1}{a}\varepsilon ^{4,0}\otimes
\gamma_4\gamma_\nu +a\varepsilon ^{0,\nu}\otimes I_4 - \varepsilon
^{4,\nu}\otimes \gamma_4\right) .
\]
One may verify that the equations
\begin{equation}
\mathcal{F}_{nm }\mathcal{F}_{mi }=B_n B_i-B^2 \delta_{ni} ,~~~~
\mathcal{F}_{nm }\mathcal{F}_{mi }\mathcal{F}_{ik }=-B^2
\mathcal{F}_{nk }
 \label{26}
\end{equation}
are hold, where $B^2=B_m^2$,
$B_m=(1/2)\epsilon_{mnk}\mathcal{F}_{nk }$ is the strength of the
magnetic field. Eq. (26) allow us to obtain the relation for the
matrix $\Sigma\equiv m+\Pi K$:
\begin{equation}
\Sigma^4 -4m\Sigma^3+\left(6m^2-b \right)\Sigma^2 + 2m\left(b-2m^2
\right)\Sigma+m^4-bm^2 =0 , \label{27}
\end{equation}
where $b=a^2\kappa_1^2B^2$. From Eq. (27), we find the inverse
matrix $\Sigma^{-1}$:
\[
\Sigma^{-1}=\frac{1}{m^2\left(
b-m^2\right)}\left[\Sigma^3-4m\Sigma^2+ \left(6m^2-b \right)\Sigma
+2m\left(b-2m^2 \right)\right]
\]
\begin{equation}
 = \frac{1}{m} +\frac{1}{m^2\left(b-m^2\right)}\biggl[i\kappa_1
\left(m^2-b\right)\mathcal{F}_{m\nu }+am\kappa_1^2\mathcal{F}_{mk
}\mathcal{F}_{k\nu } \label{28}
\end{equation}
\[
-ia^2\kappa_1^3\mathcal{F}_{mk }\mathcal{F}_{kn }\mathcal{F}_{n\nu
}\biggr]\left(\varepsilon ^{m,0}\otimes \gamma_\nu+a\varepsilon
^{m,\nu}\otimes I_4 \right) .
\]


\begin{thebibliography}{99}

\bibitem{Kruglov} S. I. Kruglov, Ann. Fond. Broglie \textbf{29},
1005 (2004); Errata-ibid (in press); arXiv: quant-ph/0408056.

\bibitem{Barut} A. O. Barut, Phys. Lett. \textbf{73B}, 310 (1978); Phys. Rev. Lett.
\textbf{42}, 1251 (1979).

\bibitem{Barut1} A. O. Barut, P. Cordero, and G. C. Ghirardi, Nuovo. Cim. \textbf{A66}, 36
(1970).

\bibitem{Barut2} A. O. Barut, P. Cordero, and G. C. Ghirardi, Phys. Rev.
\textbf{182}, 1844 (1969).

\bibitem{Wilson} R. Wilson, Nucl. Phys. \textbf{B68}, 157 (1974).

\bibitem{Dvoeglazov} V. V. Dvoeglazov, Int. J. Theor. Phys.
\textbf{37}, 1009 (1998); Ann. Fond. Broglie \textbf{25}, 81
(2000); Hadronic J. \textbf{26}, 299 (2003).

\bibitem{Kruglov1} S. I. Kruglov, Quantization of fermionic fields with
two mass states in the first order formalism, to appear in the
proceedings of 18th Workshop on Hadronic Mechanics Honoring the
70th Birthday of Prof. R.M. Santilli (Karlstad, Sweden, 20-22 Jun
2005); arXiv: hep-ph/0510103.

\bibitem{Ahieser}  A. I. Ahieser, and V. B. Berestetskii, Quantum
Electrodynamics (New York: Wiley Interscience, 1969).

\bibitem{Kruglov2} S.I.Kruglov, Symmetry and Electromagnetic Interaction of
Fields with Multi-Spin (Nova Science Publishers, Huntington, New
York, 2001).

\bibitem{Foldy} L.L. Foldy, and S. A. Wouthuysen, Phys. Rev.
\textbf{78}, 29 (1950).

\end{thebibliography}
\end{document}